\documentclass[final,3p,times,letter]{elsarticle}
\usepackage{amsmath,amssymb,bm,color,graphicx,subfig,url,hyperref} 

\newcommand{\rd}{\mathrm{d}}
\newcommand{\ri}{\mathrm{i}}

\journal{Wave Motion}

\begin{document}

\begin{frontmatter}

\title{Nonlinear waves in electromigration dispersion in a capillary}

\author[icc]{Ivan~C.~Christov}
\ead{christov@purdue.edu}
\ead[url]{http://christov.tmnt-lab.org}
\address[icc]{School of Mechanical Engineering, Purdue University, West Lafayette, IN 47907, USA}


\begin{abstract}
We construct exact solutions to an unusual nonlinear advection--diffusion equation arising in the study of Taylor--Aris (also known as shear) dispersion due to electroosmotic flow during electromigration in a capillary. An exact reduction to a Darboux equation is found under a traveling-wave anzats. The equilibria of this ordinary differential equation are analyzed, showing that their stability is determined solely by the (dimensionless) wave speed without regard to any (dimensionless) physical parameters. Integral curves, connecting the appropriate equilibria of the Darboux equation that governs traveling waves, are constructed, which in turn are shown to be asymmetric kink solutions ({\it i.e.}, non-Taylor shocks). Furthermore, it is shown that the governing Darboux equation exhibits bistability, which leads to two coexisting non-negative kink solutions for (dimensionless) wave speeds greater than unity. Finally, we give some remarks on other types of traveling-wave solutions and a discussion of some approximations of the governing partial differential equation of electromigration dispersion.
\end{abstract}


\end{frontmatter}

\section{Introduction}

From the early work on Taylor cones and jetting \cite{tm69} and the Taylor--Melcher leaky dielectric model \cite{s97} to modern microfluidics applications \cite{ssa04,sq05}, such as electrophoretic separation of mixtures \cite{g06}, electrokinetic and electrohydrodynamic phenomena remain an active topic of research \cite{b15} in physicochemical hydrodynamics \cite[Chapters~6--7]{p05} (see also \cite[Chapters~8--10]{b08}). A related topic is the phenomenon of Taylor--Aris dispersion \cite{t53,a56}, which is one example of a macrotransport process \cite{be93} describing the spread of the cross-sectionally averaged concentration of a tracer in a laminar flow far downstream from the point of injection. For example, in a pressure-driven laminar flow (mean speed $\bar{u}$) of a Newtonian fluid in a planar slot of height $h_0$, a dispersivity $\mathcal{D} = D+h_0^2\bar{u}^2/(210D)$ is found for the case of a tracer of diffusivity $D$ \cite[Chapter~2]{be93}. The cross-sectionally averaged concentration is, then, advected downstream with constant speed $\bar{u}$ and spreads with an effective diffusivity equal to the dispersivity $\mathcal{D}$.

The classical dispersion equation is a linear advection-diffusion equation, which is solvable by classical techniques (integral transforms, self-similarity methods, separation of variables, etc.\ \cite{s06}). However, different physical contexts present new aspects to dispersion. For example, streamwise variations of the mean flow in a radial geometry \cite{sb99} lead to nonstandard self-similarity exponents (specifically $t^{1/4}$ instead of $t^{1/2}$), while a shear-rate dependent diffusivity turns out to yield the classical Taylor--Aris dispersion equation albeit with different numerical pre-factors \cite{cs14}. The effect of geometry has also been addressed \cite{gcb87,b06,a06} due to its implications for microfluidic devices \cite{ssa04}, in which case a surprising result arises: the dispersivity scales with the width and not the height of the channel. 

While the latter examples all involve dispersion equations that are \emph{linear}, recent work on Taylor--Aris dispersion of concentrated suspensions has yielded \emph{nonlinear} governing equations. For example, if the diffusivity is made concentration-dependent \cite{gs12} due to the effects of shear-induced migration of finite-sized particles in a laminar flow \cite{ebs77,la87}, the dispersion equation becomes an advection-diffusion equation with nonlinear diffusivity. In this case, self-similar solutions can still be found analytically \cite[Eq.~(16)]{gs12}. If a suspension-balance approach is used to fully account for the suspended particles' effect on the laminar flow field, then the dispersion equation has both nonlinear advective and nonlinear diffusive terms \cite[Eq.~(4.1)]{r13}. In the latter case, however, the form of the nonlinearity is quite involved, precluding any analytical results. Returning to the electroosmotic flow context \cite{g02,g02b,g02c}, dispersion is considered a hinderance for lab-on-a-chip technologies if the goal is separation and fractionation \cite{g06,dg09}, while the enhanced mixing due to dispersion is sought out for other applications \cite[\S3]{ssa04}. Either way, it is clear that accurate model equations and their solutions (analytical, if possible) are needed to gain practical understanding and determine conditions for minimizing/maximizing dispersion \cite{rad11}. More recently, Ghosal and Chen \cite{gc10,cg11,cg12,gc12} have undertaken the study of electromigration and dispersion of analytes with and without a background electroosmotic flow. 

In particular, in \cite{gc12}, a \emph{nonlinear} electromigration dispersion equation that models Taylor--Aris (shear) dispersion in a capillary in the presence of electroosmotic flow was derived. This equation \cite[Eq.~(3.23)]{gc12}, which features both nonlinear advective and nonlinear diffusive terms is unlike \cite[Eq.~(4.1)]{r13}, in that the nonlinearities are elementary functions of the dependent variable. Such equations are generally of interest in mechanics and applied mathematics because of what Crighton~\cite{c86} has termed, in the context of hydrodynamics, the \emph{Taylor--Lighthill balance}: the tendency of (linear or nonlinear) dissipation such as diffusion to counteract wavefront steepening due to advective nonlinearities in the governing equations~\cite{c86,l56}. The Taylor--Lighthill balance results in smooth waveforms, often termed \emph{kinks} (to be precisely defined below), in the traveling wave context \cite{cjcbwv16}. (In higher-order evolution equations, such as the celebrated Korteweg--de Vries model, this balance produces \emph{solitons} \cite[\S2.1]{dj89}.) Ghosal and Chen \cite{gc10,cg11,cg12,gc12} presented numerical evidence, and analytical results in some special cases, showing that the Taylor--Lighthill balance yields permanent traveling electromigration waveforms. However, a detailed mathematical analysis of \cite[Eq.~(3.23)]{gc12} is lacking. Additionally, in \cite{gc12}, the authors also suggested the use of a modified version of the original nonlinearities (specifically, a Taylor-series expansion keeping only a few terms) of the electromigration dispersion equation. It is known, however, that such modifications of the nonlinearities are not always valid and can lead to significant differences between the original and approximate equations in the context of nonlinear acoustics \cite{ccj07,ccj15}. Therefore, there is an impetus to obtain \emph{exact} results regarding the \emph{full} nonlinear electromigration dispersion equation.

To this end, in the present work, we show that exact traveling wave solutions can be constructed for the electromigration dispersion equation, and the latter can be well approximated by Taylor shock ({\it i.e.}, ``tanh'') \cite{t10,c86} profiles. Aside from the fundamental, mathematical interest in obtaining an exhaustive classifications of various solutions to field theories \cite{kcs15}, it is important to understand traveling electromigration wave phenomena \cite{fps88} (also known as \emph{isotachophoretic boundaries} \cite{sp86} in other contexts) because, for example, such traveling waves can be used as the basis for electrophoretic separation methods \cite{sp86}. In this context, being able to generate and propagate a kink through an ionic solution leads to the partitioning of the solution into zones of nearly uniform compositions \cite{sp86}, effectively separating analytes of different conductivities. Alternatively, a traveling-wave can be imposed through the applied electric field, which also yields an effective separation technique for microfluidic devices \cite{e09}.

\section{Problem formulation}

Ghosal and Chen \cite[Eq.~(3.23)]{gs12} derived the following macrotransport equation for the cross-sectionally averaged sample concentration  $\bar{\phi}(x,t)$ (relative to the background):
\begin{equation}
\frac{\partial\bar{\phi}}{\partial t} + \frac{\partial}{\partial x}\left[\left( u_\mathrm{eo} + \frac{v_0}{1-\alpha \bar{\phi}}\right)\bar{\phi}\right]
= \frac{\partial}{\partial x}\left\{\left[D+\frac{k u_\mathrm{eo}^2 w_0^2}{D}\left(\frac{\alpha \bar{\phi}}{1-\alpha \bar{\phi}} \right)^2\right]\frac{\partial \bar{\phi}}{\partial x}\right\} \qquad (0<\bar{\phi} < 1/\alpha),
\label{eq:gc_eo}
\end{equation}
where $t$ is time, $x$ is the streamwise coordinate, the positive constant $u_\mathrm{eo}$ is the mean electroosmotic flow speed, the positive constant $v_0$ is the migration velocity of any isolated ion, $\alpha$ is a constant parameter used in \cite{gc10} to characterize the nonlinearity, $D$ is the constant diffusivity of each ion species (all equal in this case), $w_0$ is the planar capillary's constant half-width and $k = 2/105$. 

To simplify the analysis, we rescale $\bar{\phi}$ and switch to the moving frame of the mean flow:
\begin{equation}
x^*=x-u_\mathrm{eo}t,\qquad t^*=t,\qquad \bar{\phi}^*(x^*,t^*) = \alpha\bar{\phi}(x,t),
\end{equation}
as in standard in Taylor--Aris dispersion analysis \cite{t53,a56}. Then, Eq.~\eqref{eq:gc_eo} becomes
\begin{equation}
\frac{\partial \bar{\phi}^*}{\partial t^*} + v_0\frac{\partial}{\partial x^*}\left(\frac{\bar{\phi}^*}{1-\bar{\phi}^*}\right) = D\frac{\partial}{\partial x^*}\left\{\left[1+\frac{k u_\mathrm{eo}^2w_0^2}{D^2}\left(\frac{\bar{\phi}^*}{1-\bar{\phi}^*} \right)^2\right]\frac{\partial \bar{\phi}^*}{\partial x^*}\right\} \qquad (0<\bar{\phi}^* < 1).
\label{eq:gc_eo2}
\end{equation}
Next, we introduce dimensionless variables via the set of transformations:
\begin{equation}
x^* = w_0 X,\qquad t^* = (w_0/v_0) T,\qquad \bar{\phi}^*(x^*,t^*) = \Phi(X,T).
\end{equation}
Then, Eq.~\eqref{eq:gc_eo2} becomes
\begin{equation}
\frac{\partial \Phi}{\partial T} + \frac{\partial}{\partial X}\left(\frac{\Phi}{1-\Phi}\right) = \frac{1}{Pe} \frac{\partial}{\partial X}\left\{\left[1+k u_*^2 Pe^2\left(\frac{\Phi}{1-\Phi} \right)^2\right]\frac{\partial \Phi}{\partial X}\right\} \qquad (0<\Phi< 1),
\label{eq:gc_eo3}
\end{equation}
where we have introduced the P\'eclet number $Pe = v_0 w_0/D (>0)$ \cite{p05} and set $u_* = u_\mathrm{eo}/v_0 (>0)$ \cite{gc12}. Equation~\eqref{eq:gc_eo3} is starting point for our mathematical analysis.

\section{Traveling wave analysis: Reduction to an ODE and equilibria}
\label{sec:tws_setup}

Now, we turn our attention to right-traveling waves, {\it i.e.}, $\Phi(X,T)=F(Z)$, where $Z = (X-cT) Pe$, and the non-negative constant $c$ is a dimensionless phase speed. (Right, {\it i.e.}, downstream, traveling waves are of interest because the electroosmotic flow is from left to right \cite{gc12}. However, upstream-traveling waves are also possible, as outlined in \ref{sec:app_upstream}.) Under the latter ansatz, Eq.~\eqref{eq:gc_eo3} becomes
\begin{equation}
-cF' + \left(\frac{F}{1-F}\right)' = \left\{\left[1+\kappa\left(\frac{F}{1-F} \right)^2\right]F'\right\}' \qquad (0<F< 1),
\label{eq:gc_eo_tws}
\end{equation}
where primes denotes $\rd/\rd Z$, and we have set $\kappa := k u_*^2 Pe^2 (>0)$ for convenience.
Integrating once with resect to $Z$, and determining that the constant of integration must be zero to ensure bounded waveforms, yields the ordinary differential equation (ODE):
\begin{equation}
-cF + \frac{F}{1-F} = \left[1+\kappa\left(\frac{F}{1-F} \right)^2\right]F' \qquad (0<F< 1),
\label{eq:gc_eo_tws2}
\end{equation}
which we can recast as
\begin{equation}
 [1-c(1-F)] (1-F) F = \left[(1-F)^2 + \kappa F^2\right]F' \qquad (0<F< 1).
\label{eq:gc_eo_tws3}
\end{equation}
Equation~\eqref{eq:gc_eo_tws3} can be readily identified as a special case of \emph{Darboux' equation} \cite[\S8-6]{m60}, an \emph{autonomous} first-order ODE. As noted by Jordan {\it et al.} \cite{j12}, this is a rather unusual case because \emph{Abel's equation} \cite[\S4-1,4-2]{m60} is typically the governing ODE for the traveling wave profile in hydrodynamics (see, {\it e.g.}, \cite{cjcbwv16,ctw09,cb11}).
Finally, solving for $F'$ in Eq.~\eqref{eq:gc_eo_tws3}, we obtain
\begin{equation}
F' = \mathfrak{G}(F) := \frac{[1-c(1-F)] (1-F) F}{\left[(1-F)^2 + \kappa F^2\right]} \qquad (0<F< 1).
\label{eq:gc_eo_tws4}
\end{equation}

\begin{figure}
\centering
\includegraphics[width=0.5\textwidth]{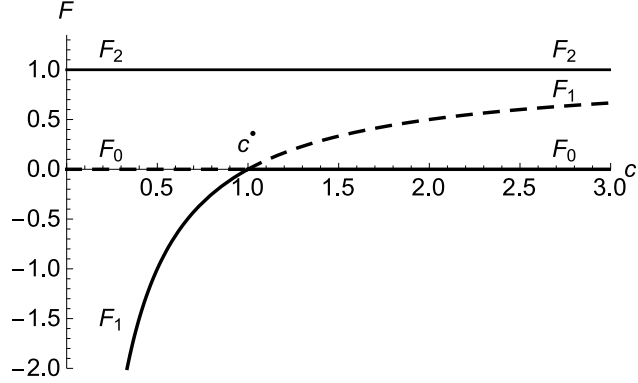}
\caption{Equilibrium solutions of Eq.~\eqref{eq:gc_eo_tws4} as functions of the traveling wave speed $c(\ge0)$. The solid and dashed curves represent stable and unstable equilibria, respectively. The bifurcation point $c=1$ is denoted by $c^\bullet$.}
\label{fig:stability_vs_c}
\end{figure}

\begin{figure}
\centering
\includegraphics[width=0.75\textwidth]{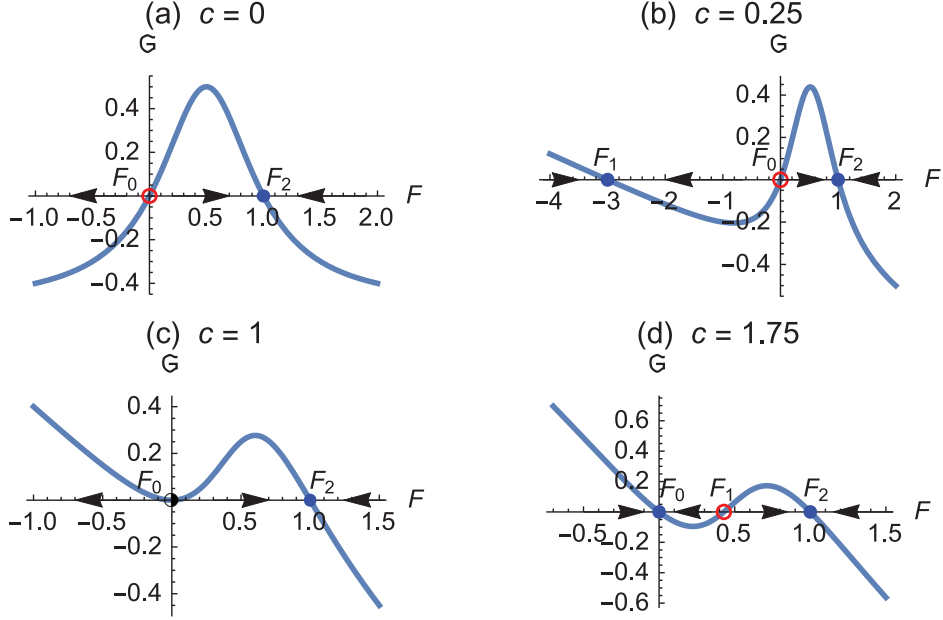}
\caption{Plot of the right-hand side $\mathfrak{G}$ of the ODE \eqref{eq:gc_eo_tws4} for three different traveling wave speeds $c$ and $\kappa=1$ (without loss of generality). In these plots,  black arrows on the $F$-axis, highlight the stability of the equilibrium points $F_{0,1,2}$, which are represented by large dots (open for unstable, filled for stable, half-filled for neutral) on the $F$-axis. Note that $F_1=\infty$ for $c=0$, therefore this equilibrium point is not visible in (a); likewise, $F_1$ and $F_0$ coalesce for $c=1$, so only one is visible in (c).}
\label{fig:stability}
\end{figure}

Next, we determine the equilibria of the ODE \eqref{eq:gc_eo_tws4} by setting $\mathfrak{G}(F)=0$ to obtain the cubic equation
\begin{equation}
F (1-F) [1-c(1-F)] = 0,
\end{equation}
which has three real solutions:
\begin{equation}
F_0 = 0,\qquad F_1 = \frac{c-1}{c} ,\qquad F_2 = 1.
\label{eq:tws_equil}
\end{equation}
For a ``subsonic'' (``speed of sound'' in this dimensionless context is $1$) traveling wave, {\it i.e.}, $c<1$, we clearly have $F_1<F_0<F_2$. Meanwhile, for a ``supersonic'' traveling wave, {\it i.e.}, $c>1$, we have $F_0<F_1<F_2$. In the special case of $c=1$, $F_0 = F_1$.
To determine the stability of the equilibria, we evaluate 
\begin{equation}
\mathfrak{G}'(F) = \frac{c \kappa (1-F^2) F^2 - c (1-F)^4+ [ (1-\kappa)F-2]F+1}{\{ [(\kappa+1)F-2]F + 1\}^2},
\end{equation}
at each of the equilibria $F=F_{0,1,2}$. Then, it can be shown that the stability of $F_{0,1,2}$ is determined by the signs of $1-c$, ${(c-1) c}/[(c-1)^2 \kappa+1]$ and $-1/\kappa$, respectively. Therefore (keeping in mind that $c\ge0$):
\begin{itemize}
\item $F_0$ is stable if $c>1$ (``supersonic''), unstable if $c<1$ (``subsonic'');
\item $F_1$ is stable if $c<1$ (``subsonic''), unstable if $c>1$ (``supersonic'');
\item $F_2$ is always stable owing to the fact that $\kappa > 0$;
\item $F_0$ and $F_1$ coalesce into a neutral equilibrium at $0$ at the bifurcation value $c^\bullet=1$.
\end{itemize}
Figure~\ref{fig:stability_vs_c} shows the stable and unstable equilibria as solid and dashed curves, respectively, as functions of $c$.
Figure~\ref{fig:stability} also illustrates these conclusions by showing $\mathfrak{G}(F)$, specifically, for (a) $c=0$, (b) $0<c<1$, (c) $c=1$ and (d) $c>1$ with $\kappa=1$ (without loss of generality).
We observe that for all $c>0$ ($c\ne c^\bullet=1$), there's \emph{two} stable equilibria, which implies \emph{bistability}.

Therefore, on the basis of these deductions, we expect that kink solutions ({\it i.e.}, bounded solutions connecting two different equilibria as $Z \to \pm \infty$ \cite[pp.~25--26]{a09}) are allowed. Specifically, 
\begin{itemize}
\item[(i)] $c=0$ [Fig.~\ref{fig:stability}(a)]: as $Z$ goes from $-\infty$ to $+\infty$, $F$ goes from $F_0$ to $F_2$, where $F_2 > F_0 = 0$, which represents a \emph{stationary} kink;
\item[(ii)] $0<c<1$ [Fig.~\ref{fig:stability}(b)]: as $Z$ goes from $-\infty$ to $+\infty$, $F$ goes from $F_0$ to $F_1$, where $F_1<F_0 = 0$, or from $F_0$ to $F_2$, where $F_2 > F_0 = 0$;
\item[(iii)] $c=1$ [Fig.~\ref{fig:stability}(c)]: as $Z$ goes from $-\infty$ to $+\infty$, $F$ goes from $F_0$ to $F_2$, where $F_2>F_0 = 0$;
\item[(iv)] $c>1$ [Fig.~\ref{fig:stability}(d)]: as $Z$ goes from $-\infty$ to $+\infty$, $F$ goes from $F_1$ to $F_0$, where $F_1 > F_0 = 0$, or from $F_1$ to $F_2$, where $F_2 > F_1$.
\end{itemize}
In all cases, the kinks are either monotonically decreasing or monotonically increasing functions of $Z$ (as required by \cite[Theorem~8.3]{asy96}). Such solutions are classified as \emph{non-Taylor shock} profiles, where by \emph{Taylor shock} we mean the classical thermoviscous hydrodynamic shock ({\it i.e.}, ``tanh'') profile \cite{t10,c86}. (Others refer to such kink solutions as \emph{dispersed shocks} \cite{l56,os69} or \emph{diffusive solitons} \cite[Chapter~11]{r99}).
There are, of course, also unbounded solutions, which are not of interest due to their lack of physical relevance. Likewise, the non-positive kink profile in case (ii) above ({\it i.e.}, the kink from $F_0$ to $F_1$ for $0< c < 1$) must also be discarded since $F$ represents a (scaled) concentration, which must be positive (specifically, $0< F< 1$ for $|Z|< \infty$).
Therefore, case (iv), in which two non-negative kink solutions ``coexist,'' is the only physically relevant manifestation of bistability in this problem.

Finally, we note that the ODE \eqref{eq:gc_eo_tws4} has no singular points and, consequently, no cancelations between factors of numerator and denominator can occur; the roots of the denominator are strictly complex: $F= (1 \pm \ri \sqrt{\kappa})^{-1}$, where recall that $\sqrt{\kappa} = \sqrt{k}\, u_* Pe > 0$. It follows that the solutions of the ODE \eqref{eq:gc_eo_tws4} will be such that $F\in C^{\infty}(\mathbb{R})$.
Next, we proceed to construct these \emph{exact} solutions.



\section{Construction of traveling wave profiles}

\subsection{Integral curves: Implicit solutions}
\label{sec:kink}

Having discarded strictly non-positive and unbounded wave profiles as unphysical, we are left with five distinct kink solutions to construct, one in cases (i) through (iii) each and two in case (iv). The kink solutions correspond to integral curves of Eq.~\eqref{eq:gc_eo_tws4} connecting different equilibrium points. To compute the integral curves, we separate variables in Eq.~\eqref{eq:gc_eo_tws4} and integrate:
\begin{equation}
\int \frac{(1-F)^2 + \kappa F^2}{ F (1-F) [1-c(1-F)]} \,\rd F = Z-Z_0,
\label{eq:tws_sep}
\end{equation}
where $Z_0$ is the constant of integration.
Then, a partial fractions expansion yields
\begin{equation}
\frac{(1-F)^2 + \kappa F^2}{ F (1-F) [1-c(1-F)]} =  \frac{(c-1)^2\kappa + 1}{(c-1) [1-c (1-F)]} - \frac{1}{(c-1) F} + \frac{\kappa}{1-F},
\end{equation}
which allows for a straightforward calculation of the integral on the left-hand side of Eq.~\eqref{eq:tws_sep}. Finally, we obtain an \emph{exact} but \emph{implicit} expression for the traveling wave profile:
\begin{equation}
\frac{\big[(c-1)^2 \kappa + 1\big] \ln [1-c (1-F)] - \big[(c-1) \kappa \ln (1-F)+\ln F\big]c}{(c-1) c} = Z-Z_0 \qquad (c\not\in\{0,1\}).
\label{eq:tws_sep2}
\end{equation}
Note that the limits $c\to0^+$ and $c\to c^\bullet = 1$ are well defined and yield
\begin{equation}
-(\kappa +1)(F-1) - \kappa  \ln (1-F) + \ln F = Z-Z_0 \qquad (c=0),
\label{eq:tws_sep2_c0}
\end{equation}
and
\begin{equation}
1- \frac{1}{F} -\kappa \ln (1-F) - \ln F = Z-Z_0 \qquad (c=c^\bullet=1),
\label{eq:tws_sep2_c1}
\end{equation}
respectively. Despite the apparent simplifications for $c=0$ and $c=1$, the implicit relations in Eqs.~\eqref{eq:tws_sep2_c0} and \eqref{eq:tws_sep2_c1} cannot be inverted to give $F(Z)$ explicitly. For $\kappa = 0$, however, Eqs.~\eqref{eq:tws_sep2_c0} and \eqref{eq:tws_sep2_c1} have explicit solutions $F(Z) = W_0\left(-e^{Z-Z_0-1}\right)$ and $F(Z) = -1/W_0\left(-e^{Z-Z_0-1}\right)$, respectively. (Here and below, $W_0$ represents the principal branch of the Lambert $W$-function \cite{c96}; see, {\it e.g.}, \cite{j14} and the references therein for a surprising number of applications of Lambert $W$ in the physical sciences.) Both of these are unbounded and, therefore, unphysical. Chen and Ghosal \cite{cg12} (see also \cite{gc10}) studied a version of the problem in which $u_*=0$ (no electroosmotic flow), hence $\kappa=0$; a related equation also arises in deionization shocks in microstructures \cite[Eq.~(15)]{mb11}. However, in the electromigration dispersion context, both $u_*$ and $Pe$ are expected to be positive \cite{gc12}.

From Eq.~\eqref{eq:tws_sep2}, we can easily obtain the asymptotics of the kink solutions near the equilibrium points in Eq.~\eqref{eq:tws_equil}. Specifically, 
\begin{equation}
F(Z) \sim \begin{cases} (1-c)^{1/[\Xi(\kappa,c)c]} \exp\left[-(c-1) (Z-Z_0)\right],\quad &F\to F_0=0,\\[3mm]
\displaystyle\frac{c-1}{c}+ \frac{1}{c}\left[\left(\displaystyle\frac{c-1}{2 c-1}\right)^{c} 2^{(c-1) (\kappa +1)} e^{(c-1) c Z}\right]^{\Xi(c,\kappa)} ,\quad &F\to F_1 = \displaystyle\frac{c-1}{c},\\[3mm]
1-\displaystyle\frac{1}{c-1}W_0\left[(c-1) e^{-(Z-Z_0)/\kappa}\right] ,\quad &F\to F_2=1,
\end{cases}\qquad (c\not\in\{0,1\}),
\label{eq:asympt_F}
\end{equation}
where we have set $\Xi(c,\kappa) := [(c-1)^2 \kappa +1]^{-1}$ for conveniences.
The $c\to0^+$ limit requires special handling, in which case we use Eq.~\eqref{eq:tws_sep2_c0} to obtain
\begin{equation}
F(Z) \sim \begin{cases} e^{-(\kappa+1) + Z - Z_0},\quad &F\to F_0=0\quad (Z\to-\infty),\\[3mm]
1 + W_0\left[-e^{-(Z-Z_0)/\kappa}\right] ,\quad &F\to F_2=1\quad (Z\to+\infty),
\end{cases}\qquad (c=0).
\end{equation}
Likewise, the $c\to c^\bullet=1$ limit must also be handled separately because $c=c^\bullet$ is a bifurcation point; in this case, we use Eq.~\eqref{eq:tws_sep2_c1} to obtain
\begin{equation}
F(Z) \sim \begin{cases} -1/W_0\left(-e^{Z-Z_0-1}\right),\quad &F\to F_0=0\quad (Z\to-\infty),\\[3mm]
1-e^{-(Z-Z_0)/\kappa} ,\quad &F\to F_2=1\quad (Z\to+\infty),
\end{cases}\qquad (c=c^\bullet=1).
\label{eq:F_asymp_c1}
\end{equation}

By translational invariance of the ODE \eqref{eq:gc_eo_tws4}, $Z_0$ is arbitrary. To fix this degree of freedom, it is customary to set $Z_0$ by requiring that (see, {\it e.g.} \cite{cjcbwv16,j12,cb11})
\begin{equation}
F(0) = \tfrac{1}{2}\big[ F(+\infty) + F(-\infty) \big],
\label{eq:F0}
\end{equation} 
where $F(\pm\infty)\equiv \lim_{Z\to\pm\infty}F(Z)$.
In particular, as shown in Section~\ref{sec:tws_setup}, the only non-negative solutions for wave speeds $c\in[0,1]$ are such that $\lim_{Z\to+\infty} F(Z) = F_2 = 1$ and $\lim_{Z\to-\infty} F(Z) = F_0=0$, hence $F(0) = 1/2$.
For $c>1$, however, there are two non-negative kinks and, therefore, $F(0) = \tfrac{1}{2}(F_1+F_0) = (c-1)/(2c)$ for the kink from $F_1$ to $F_0$ and $F(0) = \tfrac{1}{2}(F_2+F_1) = (2c-1)/(2c)$ for the kink from $F_1$ to $F_2$. Requiring that $F(0)$ be these latter values, we obtain
\begin{equation}
Z_0 = \begin{cases}
-\frac{1}{2} \left[ \kappa + 1 + 2 (\kappa -1) \log 2 \right], &c=0,\\[4mm]
\displaystyle -\frac{\big[(c-1) \kappa+1\big]c \ln 2 + \big[(c-1)^2 \kappa+1\big] \ln(1-c/2)}{(c-1) c}, &0< c< 1,\\[4mm]
1-(\kappa +1)\ln 2, &c=c^\bullet=1,\\[4mm]
\begin{cases} \displaystyle -\frac{\left[(c-1)^2 \kappa+1\right] \ln \left(\frac{1-c}{2}\right) +  \left[ (c-1) \kappa \ln \left(\frac{2 c}{c+1}\right)+\ln \left(\frac{2 c}{c-1}\right)\right]c}{(c-1) c}, &\text{kink from $F_1$ to $F_0$},\\[4mm] \displaystyle -\frac{(c-1) \left[(\kappa +1) \ln 2 + c \kappa  \ln c\right]-c \ln \left(\frac{2c-1}{c}\right)}{(c-1) c}, &\text{kink from $F_1$ to $F_2$},\end{cases}, &c>1.
\end{cases}
\end{equation}
Note that the limit as $c\to1^+$ of the value $Z_0$ for the kink from $F_1$ to $F_0$ is undefined. This is because this solution ceases to exist as the bifurcation point $c=c^\bullet =1$ is approached from above. Instead, it is the kink from $F_1$ to $F_2$ that limits onto the solution for $c=1$ and the appropriate value of $Z_0$.

Thus, we have completely determined, albeit implicitly, the exact non-negative solutions of Eq.~\eqref{eq:gc_eo_tws4}, which are the traveling wave solutions of the dimensionless electromigration dispersion equation \eqref{eq:gc_eo3}. Next, we propose some \emph{explicit} but \emph{approximate} solutions based on the results from this subsection.

\subsection{Shock thickness and a Taylor shock approximation}

In gas dynamics, the shock thickness of a kink is defined \cite[p.~590--591]{p81} as
\begin{equation}
\ell = \left|\frac{\displaystyle F(+\infty) - F(-\infty)}{F'(0)}\right|,
\label{eq:shock_thick} 
\end{equation}
where recall that $F'(0) \equiv F'(Z=0) \equiv \mathfrak{G}\big(F(Z=0)\big)$ with $\mathfrak{G}$ defined in Eq.~\eqref{eq:gc_eo_tws4} and $F(0)$ as set by Eq.~\eqref{eq:F0}. Performing the calculations for the three different cases, we obtain
\begin{equation}
\ell = \begin{cases}
\displaystyle\frac{2(1+\kappa)}{2-c}, &0\le c \le 1,\\[5mm]
\begin{cases} \displaystyle -\frac{2 \left[(c-1)^2 \kappa+(c+1)^2\right]}{(c^2-1)c}, &\text{kink from $F_1$ to $F_0$},\\[4mm] \displaystyle \frac{2 \left[(2c-1)^2 \kappa+1\right]}{(2 c-1)c}, &\text{kink from $F_1$ to $F_2$},\end{cases}, &c>1.
\end{cases}
\label{eq:shock_thick2} 
\end{equation}
Note that, now, neither the steady kink ($c=0$) nor the kink traveling at the bifurcation wave speed $c=c^\bullet=1$ require special treatment, as the value of $\ell$ in those two cases readily follows from the first case of Eq.~\eqref{eq:shock_thick2}.

If the kink solutions derived in Section~\ref{sec:kink} have a high degree of symmetry, then they will be well approximated by a symmetric Taylor shock ({\it i.e.}, $\tanh$) profile that connects the appropriate equilibria and has the appropriate shock thickness computed via Eq.~\eqref{eq:shock_thick2}. In other words,
\begin{equation}
F(Z) \approx \tfrac{1}{2}\big[F(+\infty) + F(-\infty)\big] + \tfrac{1}{2}\big[F(+\infty) - F(-\infty)\big] \tanh(2Z/\ell).
\label{eq:taylor_shock}
\end{equation}


\subsection{The kink solutions illustrated}
\label{sec:kink_examples}

\begin{figure}
\centering
\includegraphics[width=0.75\textwidth]{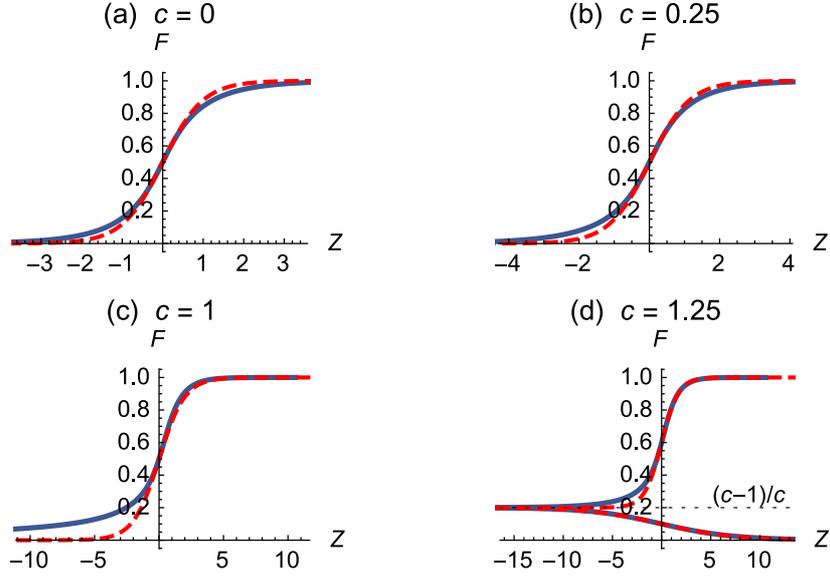}
\caption{Solid: exact kink solutions from Eqs.~\eqref{eq:tws_sep2} ($c\not\in\{0,1\}$), \eqref{eq:tws_sep2_c0} ($c=0$) and \eqref{eq:tws_sep2_c1} ($c=c^\bullet=1$). Dashed: Taylor shock approximation based on Eq.~\eqref{eq:taylor_shock} with appropriately specified limits $F(\pm\infty)$ and shock thickness $\ell$ given by Eq.~\eqref{eq:shock_thick2}. Four values of $c$ are chosen to highlight all four possible cases (i)--(iv) deduced in Section~\ref{sec:tws_setup}. Note the different horizontal scales in each plot; $\kappa=1$.}
\label{fig:kink}
\end{figure}

Based on the foregoing arguments, there are \emph{five} distinct exact kink solutions of physical interest:
\begin{itemize}
\item[(I)] $c=0$: stationary kink;
\item[(II)] $0<c<1$: subsonic kink;
\item[(III)] $c=1$: kink at the bifurcation wave speed;
\item[(IV)] $c>1$: supersonic kink from $F_0$ to $F_1$, where $F_1>F_0=0$;
\item[(V)] $c>1$: supersonic kink from $F_1$ to $F_2$, where $(0<)F_1<F_2=1$.
\end{itemize}
These five cases are illustrated in the four panels of Fig.~\ref{fig:kink} for $\kappa=\mathcal{O}(1)$ (specifically, $\kappa=1$). As can be seen from Fig.~\ref{fig:kink}, the kink profiles constructed in Section~\ref{sec:kink} as exact traveling wave solutions of the electromigration dispersion equation \eqref{eq:gc_eo3} are close to symmetric and, therefore, well approximated [for $\kappa=\mathcal{O}(1)$] by the Taylor shock profile in Eq.~\eqref{eq:taylor_shock} with appropriate values of $F(\pm\infty)$ and $\ell$. The only case in which the Taylor shock profile fails to be a good approximation is at the bifurcation value of the wave speed, {\it i.e.}, $c=c^\bullet=1$. In this case, there is strong asymmetry between the $Z=\pm\infty$ asymptotes.

\begin{figure}
\centering
\includegraphics[width=0.75\textwidth]{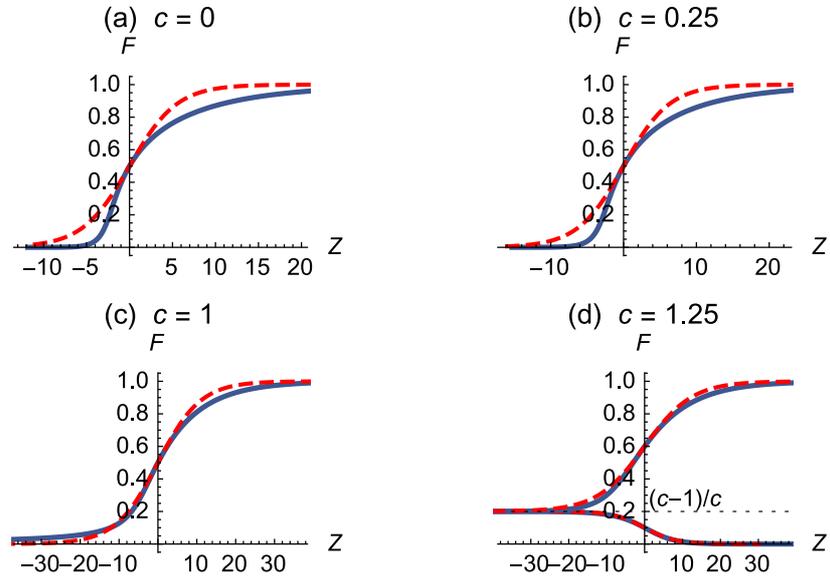}
\caption{Solid: exact kink solutions from Eqs.~\eqref{eq:tws_sep2} ($c\not\in\{0,1\}$), \eqref{eq:tws_sep2_c0} ($c=0$) and \eqref{eq:tws_sep2_c1} ($c=c^\bullet=1$). Dashed: Taylor shock approximation based on Eq.~\eqref{eq:taylor_shock} with appropriately specified limits $F(\pm\infty)$ and shock thickness $\ell$ given by Eq.~\eqref{eq:shock_thick2}. Four values of $c$ are chosen to highlight all four possible cases (i)--(iv) deduced in Section~\ref{sec:tws_setup}. Note the different horizontal scales in each plot; $\kappa=10$.}
\label{fig:kink2}
\end{figure}

Ghosal and Chen \cite[\S5]{gc12} have argued that $Pe\simeq 10$, while $u_*\simeq 1$ (under most circumstances) or $u_*\ll 1$ (for channels with low $\zeta$-potential). Therefore, $\kappa \equiv k u_*^2 Pe^2 \simeq 2$ (recall that $k = 2/105 \simeq 0.02$). In the case of a coated capillary, for which $u_* \ll 1$ say $u_*=0.1$, we would have $\kappa \equiv k u_*^2 Pe^2 \simeq 0.02$, which does not appreciably change the qualitative match between the exact kinks and their Taylor shock approximations in Fig.~\ref{fig:kink}. If, however, $\kappa$ is on the order of 100s or, even, 1000s, which could be the case at very large P\'eclet numbers (advection-dominated regime), then the kinks become markedly asymmetric in the ``subsonic'' ($c<1$) regime, as shown in Fig.~\ref{fig:kink2}.

\section{Conclusion}

In the present work, we have constructed exact solutions to an unusual nonlinear advection--diffusion equation \eqref{eq:gc_eo} arising in the study of Taylor--Aris (shear) dispersion due to electroosmotic flow during electromigration in a capillary. Specifically, under a traveling-wave ansatz, an exact reduction a Darboux ordinary differential equation \eqref{eq:gc_eo_tws3} has been achieved. We have performed a stability analysis of the equilibria of Eq.~\eqref{eq:gc_eo_tws3}, showing that their stability is determined solely by the (dimensionless) wave speed $c$ without regard to any (dimensionless) physical parameters ({\it i.e.}, $Pe$ and $u_*$). Integral curves, connecting the appropriate equilibria of the Darboux equation that governs the traveling wave solutions, have been constructed, which in turn have been shown to be asymmetric kink solutions ({\it i.e.}, non-Taylor shocks). Furthermore, it has been shown that the governing Darboux equation exhibits bistability, which leads to two coexisting non-negative kink solutions for (dimensionless) wave speeds greater than unity. 

Although, in the present work, we have only considered integral curves that connect different equilibria, the asymptotic values $F(\pm\infty)$ of the traveling wave are set by the boundary conditions in a given experiment. For example, the numerical study of Ghosal and Chen \cite{gc12} only considered traveling waves such that $F(\pm\infty)=0$. [Note that the ``triangular wave'' examples studied numerically in \cite{gc12} could be analyzed, in future work, by the matched asymptotic technique of Grundy \cite{g83} (see also \cite{lr11}).] Both of these cases, {\it i.e.}, $F(\pm\infty)=0$ and $F(+\infty)\ne F(-\infty)$, are relevant to Taylor--Aris dispersion. The former is special case ``(B1)'' in G.I.\ Taylor's classic paper \cite[\S5]{t53}, while the latter is special case ``(B2).'' Physically, cases (B1) and (B2) correspond to the release of a concentrated bolus at $x=0$ and $t=0$, and dissolved material at some nonzero $\Phi$ allowed to enter the capillary (initially at $\Phi=0$ for $x\ge0$) at a uniform rate at $x=0$ and $t\ge 0$, respectively. The exact traveling wave results herein apply to case (B2) and, on the basis of the principle of \emph{intermediate asymptotics} \cite{b96}, our results provide the \emph{exact} post-transient solution to the electromigration dispersion problem.

It is also noteworthy that previous studies of electromigration in the absence of electroosmotic flow ({\it i.e.}, $u_\mathrm{eo}=0\Rightarrow u_*=0\Rightarrow \kappa=0$) \cite{gc10,cg12} have explored kink solutions, such as those constructed herein. It is straightforward to specialize the results from Section~\ref{sec:kink} to the case $\kappa=0$. Note, however, that because $F_2=\infty$ for $u_*=0$, bounded kink solutions only exists in case (IV) from Section~\ref{sec:kink_examples}, {\it i.e.}, the kink from $F_0$ to $F_1$ in case (iv) of Section~\ref{sec:tws_setup}. The most conspicuous feature of taking $\kappa =0$ is that solutions with \emph{corners} are possible, {\it i.e.}, while $F\in C(\mathbb{R})$, $F'\not\in C(\mathbb{R})$. Such solutions are known as \emph{acceleration waves} in hydrodynamics \cite{j06}. The piecewise-defined acceleration wave solutions can be constructed so that they are bounded, {\it e.g.}, following \cite{j06}. Although the construction in \cite{j06} is for a different ODE, the approach can be generalized to construct acceleration waves for electromigration without dispersion ({\it i.e.}, $\kappa=0$). This, however, is beyond the scope of the present work.

In future work, it would be of significant interest to understand the physical implications of the bistability manifested by the coexisting kinks [cases (IV) and (V)] in Section~\ref{sec:kink_examples}. Likewise, it should be determined whether the \emph{upstream}-traveling waves in \ref{sec:app_upstream} can be observed in experiments. Finally, \emph{in contrast to} related nonlinear advection-diffusion equations such as the  Burgers equation \cite[\S4.3]{w99} and similar models arising in acoustics \cite{j12}, in which a unique traveling-wave speed $c$ can be determined by enforcing the limiting values $F(\pm\infty)$ of a kink, the singularity in the associated ODE \eqref{eq:gc_eo_tws4} as $F\to 1$ leaves the speed of electromigration dispersion waves arbitrary.

\section*{Acknowledgements}
The author is indebted to Dr. Pedro M.\ Jordan for extensive and helpful remarks on the Darboux equation and traveling-wave analysis, and to Prof.\ Howard A.\ Stone for many insightful discussions on Taylor--Aris dispersion.

\bibliographystyle{elsarticle-num}
\bibliography{electromigration}

\appendix
\gdef\thefigure{\arabic{figure}}%

\section{Comments on an approximate version of Eq.~\eqref{eq:gc_eo3}}

Although nonlinear diffusion and nonlinear advection-diffusion partial differential equations (PDEs) such as Eq.~\eqref{eq:gc_eo3} can be directly ``attacked'' with any number of reliable numerical methods, from Crank--Nicolson schemes with internal iterations \cite{cd02,zcs14} to Fourier--Galerkin methods \cite{cc02,cc05,b01} and central finite-volume (Godunov-type) schemes \cite{kt00,zfccs15}, Ghosal and Chen \cite[Eq.~(3.24)]{gc12} introduced an approximate version of Eq.~\eqref{eq:gc_eo3} by expanding the nonlinearities in a Taylor series for $\Phi\ll1$. Using the notation introduced above, they arrived at
\begin{equation}
\frac{\partial \Phi}{\partial T} + \frac{\partial}{\partial X}\left[\Phi(1+\Phi+\Phi^2)\right] = \frac{1}{Pe} \frac{\partial}{\partial X}\left[\left(1+\kappa^2 \Phi^2 \right)\frac{\partial \Phi}{\partial X}\right] + \mathcal{O}(\Phi^4) \qquad (0<\Phi< 1).
\label{eq:gc_approx}
\end{equation}

In general, as argued in the acoustics context in \cite{ccj07,ccj15}, such approximations can have profoundly unpredictable effects on the nature of solutions, even if the assumptions underlying the Taylor-series expansion are satisfied. Jordan {\it et al.}~\cite{j12} examined some of the latter implications on traveling waves, showing, once again, unexpectedly nontrivial differences between the traveling wave solutions of the original and the approximate equations. Here, we follow the approach in \cite{j12} and find the associated ODE of \eqref{eq:gc_approx} to be 
\begin{equation}
(1-c) F + F^2 + F^3 = \left[1+\kappa F^2\right]F' \qquad (0<F< 1),
\label{eq:gc_eo_tws_approx}
\end{equation}
after one integration and after determining the constant of integration to be zero. From Eq.~\eqref{eq:gc_eo_tws_approx}, we see that the equilibria are now
\begin{equation}
F_0 = 0,\qquad F_1 = \frac{1}{2} \left( -1 - \sqrt{4 c-3} \,\right),\qquad F_2 = \frac{1}{2} \left( -1 + \sqrt{4 c-3} \,\right),
\label{eq:tws_equil_approx}
\end{equation}
and, whereas Eq.~\eqref{eq:gc_eo_tws3} was valid $\forall c\ge0$, Eq.~\eqref{eq:gc_eo_tws_approx} requires the physically unclear restriction that $c \ge 3/4$. Moreover, it is no longer possible to even recover the equilibria of Eq.~\eqref{eq:gc_eo_tws3}, given in Eq.~\eqref{eq:tws_equil}, because $F_{1,2}$ in Eq.~\eqref{eq:tws_equil_approx} \emph{do not} have a real-valued Taylor expansion for $c\ll1$, while for $|c-3/4|\ll1$, the Taylor expansion is real-valued but does not yield the equilibria in Eq.~\eqref{eq:tws_equil}. Apparently, the approximation of Eq.~\eqref{eq:gc_eo3} by Eq.~\eqref{eq:gc_approx}, while well-intentioned, is ultimately unfruitful.
Exploring all the differences between these two equations is beyond the scope of the present work.


\section{Upstream waves ($c<0$)}
\label{sec:app_upstream}

\begin{figure}
\centering
\includegraphics[width=0.75\textwidth]{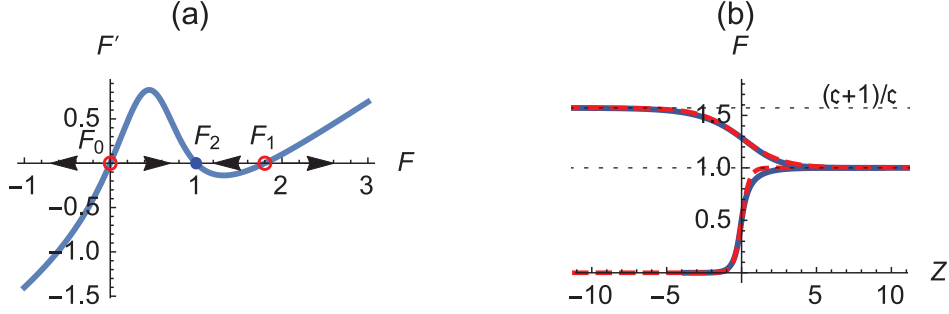}
\caption{(a) Plot of $F'$ versus $F$ for the ODE \eqref{eq:gc_eo_tws3_l} governing left- ({\it i.e.}, upstream-) traveling waves with $\mathfrak{c}=1.25$ and $\kappa=1$ (without loss of generality). In this plot, black arrows on the $F$-axis, highlight the stability of the equilibrium points $F_{0,1,2}$, which are represented by dots (open for unstable, filled for stable)  on the $F$-axis. (b) Solid: exact upstream-propagating kink solutions from Eqs.~\eqref{eq:tws_sep2_l}. Dashed: Taylor shock approximation based on Eq.~\eqref{eq:taylor_shock} with appropriately specified limits $F(\pm\infty)$ and shock thickness $\ell$ given by Eq.~\eqref{eq:shock_thick2_l}; $\mathfrak{c}=1.25$ and $\kappa=1$ as in (a).}
\label{fig:upstream}
\end{figure}

To describe left- ({\it i.e.}, upstream-) traveling waves, let $c=-\mathfrak{c}$, where $\mathfrak{c}>0$. Then, Eq.~\eqref{eq:gc_eo_tws3} becomes
\begin{equation}
 [1+\mathfrak{c}(1-F)] (1-F) F = \left[(1-F)^2 + \kappa F^2\right]F' \qquad (0<F< 1),
\label{eq:gc_eo_tws3_l}
\end{equation}
and its equilibria are
\begin{equation}
F_0 = 0,\qquad F_1 = \frac{\mathfrak{c}+1}{\mathfrak{c}} ,\qquad F_2 = 1.
\label{eq:tws_equil_l}
\end{equation}
Their stability is determined by the signs of $1+\mathfrak{c}$, ${(\mathfrak{c}+1) \mathfrak{c}}/[(\mathfrak{c}+1)^2 \kappa+1]$ and $-1/\kappa$, respectively. Clearly, for $c<0$, \emph{no} bifurcation occurs at $c=-c^\bullet=-1$, there no longer exists a critical wave speed and the phase portraits are qualitatively similar for all $\mathfrak{c}$ as shown in Fig.~\ref{fig:upstream}(a).  Unlike the situation in Section~\ref{sec:tws_setup} for $c\ge0$, now $F_0$ and $F_1$ are \emph{always} unstable, while $F_2$ is still always stable owing to $\kappa>0$.

Consequently, bistability persists for all $\mathfrak{c}$, and two kink solutions (one connecting $F_0$ to $F_2$ and another connecting $F_1$ to $F_2$ as $Z$ goes from $-\infty$ to $+\infty$) always coexists. From Eq.~\eqref{eq:tws_sep2}, these two kinks are implicitly given by
\begin{equation}
\frac{\big[(\mathfrak{c}+1)^2 \kappa + 1\big] \ln [1+\mathfrak{c} (1-F)] + \big[\ln F-(\mathfrak{c}+1) \kappa \ln (1-F)\big]\mathfrak{c}}{(\mathfrak{c}+1) \mathfrak{c}} = Z-Z_0 \qquad (\forall \mathfrak{c} > 0),
\label{eq:tws_sep2_l}
\end{equation}
and, from Eq.~\eqref{eq:shock_thick2}, their shock thicknesses are 
\begin{equation}
\ell = 
\begin{cases} \displaystyle \frac{2(1+ \kappa)}{2+\mathfrak{c}},\quad &\text{kink from $F_0$ to $F_2$},\\[4mm] \displaystyle \frac{2 \left[(2\mathfrak{c}+1)^2 \kappa+1\right]}{(2 \mathfrak{c}+1)\mathfrak{c}},\quad &\text{kink from $F_1$ to $F_2$},
\end{cases}\qquad (\forall \mathfrak{c} > 0).
\label{eq:shock_thick2_l} 
\end{equation}
The asymptotic behaviors of $F$ near the equilibria in Eq.~\eqref{eq:tws_equil_l} can be obtained directly from Eq.~\eqref{eq:asympt_F}.
Figure~\ref{fig:upstream}(b) highlights the structure of the coexisting upstream-traveling kinks. Note their high degree of symmetry results in a very close match to the Taylor shock approximation.

\end{document}